\documentclass[aps,letterpaper,twocolumn,prl,footinbib,floatfix,showpacs]{revtex4}
\usepackage{amsmath}
\usepackage{amsfonts}
\usepackage{amssymb}
\usepackage[scanall]{psfrag}
\usepackage{psfrag}
\usepackage{graphicx}
\usepackage{color}
\usepackage{subfigure}
\usepackage{bbold}

\usepackage{yypreamble}

\newcommand{\round}[1]{\left[\left[{#1}\right]\right]}

\begin{document}

\author{Yariv Yanay}
\affiliation{Laboratory of Atomic and Solid State Physics, Cornell University, Ithaca NY 14850}
\author{Erich J. Mueller}
\affiliation{Laboratory of Atomic and Solid State Physics, Cornell University, Ithaca NY 14850}
\title{Saving the Coherent State Path Integral}
\date{\today}

\begin{abstract}
By returning to the underlying discrete time formalism, we relate spurious results in coherent state semiclassical path integral calculations to the high frequency structure of their propagators. We show how to modify the standard expressions for thermodynamic quantities to yield correct results. These expressions are relevant to a broad range of physical problems, from the thermodynamics of Bose lattice gases to the dynamics of spin systems.
\end{abstract}

%03.65.Db 	Functional analytical methods 
%03.65.Sq 	Semiclassical theories and applications 
%03.75.-b		Matter waves
%05.30.Jp 	Boson systems
%67.85.Hj 	Bose-Einstein condensates in optical potentials 
%67.85.Fg	Multicomponent condensates; spinor condensates

\maketitle

Path integrals convert the difficult problem of diagonalizing a Hamiltonian into the potentially simpler one of summing over a set of all possible paths, weighted by the classical action \cite{Kleinert2009,Klauder2003}. They are particularly powerful for making semiclassical approximations, where only a few classical paths dominate. Often the natural variables for describing the path are conjugate. For example, one would like to describe a spin system in terms of paths on the Bloch sphere, even though the different components of spin do not commute \cite{Klauder1985}. Coherent states are often used in such cases, and can yield useful results 
\cite{Langer1968,Solari1987,Zhang1990,Kochetov1997,Zhang1999,Altland2010,Kochetov1995,Stone2000,Pletyukhov2002,Kleinert2013}. 

Here, we analyze the structure of such path integrals, demonstrating a practical scheme for eliminating anomalies which were first confronted in the 1980s \cite{Funahashi1995,Enz1986,Funahashi1995a,Belinicher1999,Baranger2001,Shibata2001,Garg2003,Viscondi2011}. The issues we address were most clearly described by Wilson and Galitski \cite{Wilson2011}, who used two simple examples to illustrate the anomalies. The particular problems described in their paper arise in the continuous-time formulation of the path integral, and we seek to correct them by returning to the discrete-time formalism. To do so, we must restrict ourselves to the semiclassical path integral, expanding the action in quadratic quantum fluctuations around a classical path. Braun and Garg \cite{Braun2007,Braun2007a} calculated the exact propagator for the discrete semiclassical path integral for the particular case of the harmonic oscillator coherent state. We perform a closely related expansion which allows for the use of a more general basis. We also present our results as a correction to the commonly used continuous-time result, providing systematics corrections to previously calculations.

One example considered in \cite{Wilson2011} is a path integral calculation of the partition function $Z_{ss}\pr$ of the single site Bose Hubbard model, $\hat H_{ss} = \frac{U}{2}\hat n\p{\hat n - 1} - \mu \hat n$, where $\hat n = \hat a\dg\hat a$ represents the number of Bosons, $U$ parameterizes their interaction and $\mu$ is the chemical potential. This is a sufficiently simple problem that one can calculate the exact partition function $Z_{ss}$, and find $Z_{ss}\ne Z_{ss}\pr$. In particular, at zero temperature, the mean occupation number calculated from $Z_{ss}\pr$ is $\avg{n\pr} = \round{\frac{\mu}{U}}$, which the exact result derived from $Z_{ss}$ is $\avg{n} = \round{\frac{\mu}{U} + \half}$. Here $\round{x}$ is the integer closest to $x$.

We derive an algorithm for correcting the path integral result for the free energy $F = -\frac{1}{\gb}\log Z$,
\begin{equation} \begin{split}
F = F^{CPI} -i \frac{1}{4\Delta t}\intrml{d\chi}{0}{\pi}e^{i\chi}
\log\br{\frac{\det G^{-1}_{\omega}}{\det \bar G^{-1}_{\omega}}}_{\omega=\frac{\pi e^{i\chi}}{\Delta t}}
\label{eq:Fcorrection}
\end{split} \end{equation}
Here $F^{CPI}$ is the free energy obtained from the continuous-time path integral (CPI)
while the matrices $\br{G_{\omega}}_{ij} = \avg{\psi^{i}_{\omega}\psi^{j}_{-\omega}}, \br{\bar G_{\omega}}_{ij} = \avg{\psi^{i}_{\omega}\psi^{j}_{-\omega}}_{CPI}$ are composed of perturbation field propagators in frequency space for a discrete-time and CPI calculation, respectively.
We precisely define all these terms below as we derive Eq.~(\ref{eq:Fcorrection}) and discuss techniques for calculating the correction terms.

As emphasized by Wilson and Galitski, our corrections are not related to ambiguities of operator ordering or geometric phases. Rather, they arise from the over-completeness of coherent states. 

The formulation of partition function as a path integral in imaginary time involves the expansion
\begin{equation} \begin{split}
Z & = \Tr e^{-\gb \hat H} = \sum_{\vec\Psi_{0}}\bra{\vec\Psi_{0}} e^{-\gb \hat H} \ket{\vec\Psi_{0}}
\\ & = \sum_{\vec\Psi_{1},\dotsc,\vec\Psi_{N_{t}}}\prod_{t=1}^{N_{t}} \bra{\vec\Psi_{t-1}}e^{-\hat H\Delta t} \ket{\vec\Psi_{t}}.
\end{split} \end{equation}
Here $\gb = 1/T$ is the inverse temperature. $\acom{\ket{\vec\Psi_{t}}}$ is any complete basis of the states, characterized by a set of parameters $\vec\Psi_{t}$, e.g. $\vec \Psi_{t} = \mat{n, \varphi}$ so that $\hat a \ket{\mat{n, \varphi}} = \sqrt{n}e^{i\varphi}\ket{\mat{n, \varphi}}$ in the coherent state basis of the Bose-Hubbard model. The sum $\sum_{\vec\Psi_{t}} \ket{\vec\Psi_{t}}\bra{\vec\Psi_{t}} = \mathcal I$ is the identity operator, of which we insert $N_{t} - 1\equiv \gb/\Delta t -1$ copies into the operator. We are now summing over all $N_{t}$-point paths in $\vec \Psi$-space, with $\vec\Psi_{0} = \vec\Psi_{N_{t}}$.
In the limit of small $\Delta t$ one can approximate $e^{-\hat H\Delta t} \approx 1 - \hat H\Delta t$ and thus write the partition function in the form of a discrete time path integral $Z = \intrm{\D \vec\Psi}e^{-\sum_{t}L_{t}}$, where the Lagrangian is
\begin{equation} \begin{split}
L_{t}  = -\log\br{\braket{\vec\Psi_{t}}{\vec\Psi_{t+1}}}& + \Delta t\frac{\bra{\vec\Psi_{t}} \hat H \ket{\vec\Psi_{t+1}}}{\braket{\vec\Psi_{t}}{\vec\Psi_{t+1}}}.
\label{eq:LEvalid}
\end{split} \end{equation}
When the basis $\acom{\ket{\vec\Psi}}$ is orthogonal, the first term in this expansion can be taken to be arbitrarily small, and one can approximate $\ket{\vec\Psi_{t+1}} \approx \p{1 + \Delta t \partial_{t}}\ket{\vec\Psi_{t}}$, and by taking $\Delta t\to 0$ convert the problem into the traditional CPI form \cite{Altland2010}. This approximation breaks down when expanding in an overcomplete basis, if the overlap between consecutive time steps remains finite for states that differ to a non-infinitesimal degree. 

As was previously noted \cite{Belinicher1999}, even in the face of this problem, the discrete time formulation in Eq.~(\ref{eq:LEvalid}) remains valid. Our task is to develop a techniques for calculations using the discrete time path integrals, and to relate them to the more familiar continuous case. In particular we wish to find a correction of the form Eq.~(\ref{eq:Fcorrection}).

To do so we follow standard procedure \cite{Smirnov2010} and characterize the states in terms of a saddle point solution $\bar{\vec\Psi}$ satisfying $\br{\frac{\gd L_{t}}{\gd\vec\Psi_{t}}}_{\vec\Psi_{t} = \bar{\vec\Psi}} = 0$, and a fluctuation $\vec\psi_{t}$, writing $\vec\Psi_{t} = \bar{\vec\Psi} + \vec\psi_{t}$. We then expand to quadratic order in the fluctuations  $L_{t} = L_{0} + \vec \psi_{t} \cdot L_{2}\cdot \vec\psi_{t} + \vec \psi_{t}\cdot L_{2\Delta}\cdot \vec\psi_{t+1} + O\p{\abs{\psi_{t}}}^{3}$ where the classical energy $L_{0}$ and matrices $L_{2}, L_{2\Delta}$ are independent of time. This saddle point approximation becomes exact as the number of local degrees of freedom become large.  For example, in the Bose Hubbard Model, it is the leading correction in a $1/n$ expansion, where $n$ is the average number of particles per site.  Similarly, in a spin system, the total spin $S$ plays the role of $n$. In terms of the Fourier components $\vec\psi_{\omega} = \frac{1}{\sqrt{N_{t}}}\sum_{t}e^{-i\omega t}\vec\psi_{t}$, the partition function reads
\begin{equation} \begin{split}
Z = \intrm{\D\psi} \exp\br{-\gb F_{0} - \half \sum_{\omega=\omega_{n}}\vec\psi_{\omega}\cdot G_{\omega}^{-1}\cdot \vec\psi_{\omega}}
\end{split} \end{equation}
where summation is over the frequencies $\omega_{n} = \frac{2\pi}{\gb} n$ for $n = -\frac{N_{t}-1}{2}\dotsc \frac{N_{t}-1}{2}$, yielding the free energy
\begin{equation} \begin{split}
F =  F_{0} + \frac{1}{\gb}\sum_{n = -\frac{N_{t}-1}{2}}^{\frac{N_{t}-1}{2}} \half\log\br{\frac{\det G_{\omega_{n}}^{-1}}{2\pi}}.
\label{eq:F}
\end{split} \end{equation}
This compares with the free energy given by the continuous-time formalism,
$F^{CPI} = F_{0}^{CPI} + \frac{1}{\gb}\sum_{n = -\infty}^{\infty} \half\log\br{\frac{\det \bar G_{\omega_{n}}^{-1}}{2\pi}} $
where $\bar G_{\omega_{n}}^{-1}$ is the CPI fluctuation matrix. As we take $\Delta t\to 0$, generically we expect the classical free energy to converge to the continuous result $F_{0} \to F_{0}^{CPI}$, and the sum $\sum_{\abs{n} >\frac{N_{t}-1}{2}} \half\log\br{\frac{\det \bar G_{\omega_{n}}^{-1}}{2\pi}} \to 0$.

The difference in energies is given then by 
\begin{equation} \begin{split}
F - F^{CPI} = \frac{1}{\gb}\sum_{n = -\frac{N_{t}-1}{2}}^{\frac{N_{t}-1}{2}} 
\half\log\br{\frac{\det G_{\omega_{n}}^{-1}}{\det \bar G_{\omega_{n}}^{-1}}}.
\label{eq:dF}
\end{split} \end{equation}
We can replace this sum with a contour integral, using the identity
\begin{equation} \begin{split}
\frac{1}{2\pi} &\oint_{\gamma} \mathrm{d\omega} \frac{f\p{\omega}}{e^{i\gb \omega} - 1} 
\\ & = \frac{1}{\gb}\sum_{\omega=\omega_{n}} f\p{\omega} + i\sum_{\omega_{f}}\text{Res}\br{\frac{f\p{\omega}}{e^{i\gb \omega} - 1},\omega_{f}}.
\label{eq:sumrule}
\end{split} \end{equation}
Here the last sum is over the poles $\omega_{f}$ of $f\p{\omega}$ inside the contour $\gamma$, and $\gamma$ is the complex circle defined by $\abs{\omega} = \frac{2\pi}{\gb}\frac{N_{t}}{2} = \frac{\pi}{\Delta t}$. The notation $\text{Res}\br{f\p{\omega},\omega_{f}}$ refers to the residue of $f\p{\omega}$ at $\omega = \omega_{f}$ and here $f\p{\omega} = \half\log\br{\frac{\det G_{\omega}^{-1}}{\det \bar G_{\omega}^{-1}}}$. 

In the present case the last term of Eq.~(\ref{eq:sumrule}) vanishes: for any fixed $\omega$, $\lim_{\Delta t\to 0}G_{\omega}^{-1} = \bar G_{\omega}^{-1}$. Thus the function $f\p{\omega}$ is analytic inside $\gamma$, and the set $\acom{\omega_{f}}$ of singularities is empty. For $\abs{\omega\Delta t}>\pi$, the matrices $G_{\omega}^{-1}$ and $\bar G_{\omega}^{-1}$ are no longer simply related, and $f\p{\omega}$ has branch cut singularities outside of $\gamma$.

Once the residue term is eliminated, we are left with the contour integral. This integral involves fluctuations of frequency $\omega_{\max} = \frac{\pi}{\Delta t}$, corresponding to the time scale separating consecutive time steps. When the basis $\ket{\vec\Psi_{t}}$ is orthogonal these fluctuations are vanishingly small, but for an overcomplete basis they are finite, and the contour integral does not vanish. Straightforward algebra then reduces Eqs.~(\ref{eq:dF}) and (\ref{eq:sumrule}) to the expression in Eq.~(\ref{eq:Fcorrection}).

A clear example of this calculation is provided by the single-site Bose-Hubbard Hamiltonian. Using the coherent state basis and the field $\vec\psi_{t} = \mat{\gd n_{t}, \phi_{t}}$, the components of the  quadratic Lagrangian are
\begin{equation} \begin{split}
& \qquad \qquad L_{0}  = \half \frac{\mu^{2}}{U}\Delta t 
\\ L_{2} & = \mat{\frac{U}{4\mu}\p{1 + \mu \Delta t} & 0 \\ 0 & \frac{\mu}{U}\p{1 - \mu \Delta t}} 
\\ & L_{2\Delta}  = -\br{1 - \mu\Delta t} \mat{\frac{U}{4\mu} & \half[i] \\ -\half[i] & \frac{\mu}{U}}
\end{split} \end{equation}
and so
\begin{equation} \begin{split}
\det G_{\omega}^{-1} = 2\p{1 - \cos\p{\omega\Delta t}}\p{1 - \mu \Delta t}.
\end{split} \end{equation}
This compares with the CPI result  $ \det \bar G_{\omega}^{-1} = \p{\gb\omega}^{2}$, and indeed the ratio of the two is finite everywhere for $\abs{\omega} \le \pi/\Delta t$. By performing the contour integral one finds the difference between the free energies  $F - F^{CPI} = -\frac{\mu}{2}$ up to an irrelevant constant.

The power of this approach is more readily apparent in the multisite Bose Hubbard model \cite{Yanay2012a}. Consider a $D$-dimensional cubic lattice of $N_{s}$ sites with lattice constant $a_{0}$. There momentum is a good quantum number and one can consider $G_{\omega,\vk}$. The large $\omega$ structure takes on the simple form
\begin{equation} \begin{split}
\frac{\det G_{\omega,\vk}^{-1}}{\det \bar G_{\omega,\vk}^{-1}} = \frac{2\p{1 - \cos\p{\omega\Delta t}}\p{1 + \gep_{k} \Delta t}}{\gb^{2}\omega^{2}}
\end{split} \end{equation}
where $\gep_{k} = 4J\sum_{j=1}^{D}\sin^{2}\p{k_{j}a_{0}/2} - \mu$. By performing the contour integral one finds simply, 
\begin{equation} \begin{split}
F - F^{CPI} = \half\p{\mu - 2J\times D}N_{s}
\end{split} \end{equation}
plus a constant. This is the same $\mu$ dependence as the single-site problem.

For completeness sake, we present the second system explored by Wilson and Galitski in \cite{Wilson2011}. We examine the Hamiltonian $\hat H = \hat S_{z}^{2}$ for a spin $S$ system. The difference in free energies between the exactly-calculated and the CPI results is given, at $T\to 0 $, by $\Delta F = -\frac{S}{2}$. Using the semiclassical formalism presented here, one finds
\begin{equation} \begin{split}
\frac{ \det G^{-1}_{\omega} }{ \det\bar G^{-1}_{\omega}} = 
\frac{2\p{1-\cos\p{\omega\Delta t}}\p{1 - \p{S - \half} \Delta t}}{\gb^{2}\omega^{2}}
\end{split} \end{equation}
leading to a correction of $F = F^{CPI} - \p{\frac{S}{2} - \frac{1}{4}}$. Our finite time-step correction accounts for most of the discrepancy, while the remaining $O\p{S}^{0}$ term arises from the semiclassical approximation.

\section{Acknowledgements}

This paper is based upon work supported by the National Science Foundation under Grant No. PHY-1068165.

\bibliography{/Users/yarivyanay/Documents/University/Citations/library}
\bibliographystyle{apsrev}

\end{document}